\documentstyle[11pt,aaspp4]{article}
\def\beq{\begin{equation}}
\def\eeq{\end{equation}}
\def\bey{\begin{eqnarray}}
\def\eey{\end{eqnarray}}
\input epsf

\begin{document}

\title
	{Reddening of microlensed LMC stars vs. the location of the lenses}
\author
	{HongSheng Zhao
	\\Sterrewacht Leiden, 
Niels Bohrweg 2, 2333 CA, Leiden, The Netherlands (hsz@strw.LeidenUniv.nl)}
\date{Accepted ........      Received .......;      in original form .......}
\label{firstpage}

\begin{abstract}
We propose an observational test that can break the degeneracy of two
main classes of microlensing models to the Magellanic Clouds: (a) the
lenses are located in the Galactic halo, and (b) the lenses are
located in the LMC disk.  The source stars in the latter
(self-lensing) models tend to be at the far side or behind the LMC
disk, thus experience more reddening and extinction by dust in the LMC
disk than ordinary stars in a nearby line of sight.  Clearly such bias
would not occur in the macho halo lensing models.  We show that this
reddening effect is at a level readily observable for the present 30
or so microlensing alerts fields, either with multi-band photometry
from a good seeing site, or more definitively with ultra-violet (UV)
spectroscopy with Space Telescope Imaging Spectrograph (STIS).  Stars
behind the LMC dust layer should stand out as UV-faint objects (by
more than 1 mag than average stars in the LMC).  HST can also resolve
numerous faint neighbouring stars within a few arcsecs of a lensed
source, hence remove blending in these crowded regions and build a
reddening map to control the patchiness of dust.
\end{abstract}

\keywords{dust, extinction -- Magellanic Clouds --- Galaxy: structure
}

\section{Introduction}

One of the main puzzles of Galactic microlensing surveys is the poorly
determined location of the lens population of the events towards the
Magellanic Clouds.  Currently there are two popular views on the
issue: (a) the lenses are located in the halo, hence are likely
baryonic dark matter candidates (Alcock et al. 1997); (b) both the
lenses and sources are part of the Magellanic Clouds, hence are stars
orbiting in the potential well of the Clouds (Sahu 1994, Wu 1994, Zhao
1998a,b, 1999a,b, Weinberg 1999).  The amount of star-star lensing is
sensitive to assumptions of the structure and equilibrium of the
Magellanic Clouds (Gould 1995, Zhao 1998a, Aubourg et al. 1999, Salati
et al. 1999, Gyuk \& Gates 1999, Gyuk, Dalal \& Griest 1999, Evans \&
Kerins 1999).   For star-star self-lensing in the LMC to be efficient,
the LMC should be fairly thick in the line of sight.  To break the
degeneracy of the models, we should design observations which are
sensitive to the location of the lens and the thickness of the LMC.
Several lines of attack have been proposed in Zhao (1999a).  For
example, a direct signature of self-lensing of a dense, but extended
stellar component, is that the lensed stars should be at the far side
of the component, hence somewhat fainter than the unlensed ones.
There are tantalizing evidences for this distance effect playing a
role both in the events towards the Galactic bulge/bar, particular two
clump giant events OGLE-BLG-3 and OGLE-BLG-10 (Stanek 1995), and in
the events towards the LMC, particularly the clump giant event
MACHO-LMC-1 (cf. Zhao et al. 1999).  Unlike the end-on cigar-shaped
Galactic bar, the LMC is an irregular disk galaxy and it is close to
face-on, so its front-to-back thickness is hard to resolve with
photometric or trigonometric parallax if the LMC is indeed thin: a
$500$pc spread in the line of sight translates to $\sim 0.02$mag in
distance modulus, or $0.2$ micro arcsec in parallax.

Here we propose a more practical test for the above two
popular models of the location of the lenses.
We propose to measure the distribution of the reddening of
individual LMC stars in small patchs of sky centered on the
microlensed stars.  Basicly some kind of ``reddening parallaxes'' can
be derived for these stars from the line of sight depth effect, i.e.,
the dust layer in the LMC makes stars behind the layer systematicly
redder than those in front of the layer.  This is a variation of the
well-known technique of differentiating the near/far side and the
trailing/leading of a spiral arm with the dust lane that runs across a
close-to-edge-on spiral galaxy.  Our method involves obtaining
multi-band photometry and/or spectroscopy of fairly faint ($19-21$mag)
stars during or well after microlensing.

After a brief account of reddening in the LMC \S2, we describe our
basic argument about the excess reddening of microlensed sources in \S3.  We
discuss several complications of the method (e.g. patchiness of dust)
in \S4.  We model the dependence on the thickness of the dust layer
in \S5.  We summarize the results and the observational strategy in \S6.

\section{Dust layer of the LMC and the Galactic foreground}
The internal extinction in the LMC is fairly small because of its
close to face-on geometry and it is patchy.  Hence it is a subtle
effect that we propose to measure.  Internal extinction of the LMC has
been studied many times in the past (e.g., Hill et al. 1994, 
Oestreicher \& Schmidt-Kaler 1996).
Harris et al. (1997) select a sample of a few thousand OB stars from
their LMC UBVI multi-band photometric survey,
and they map out dust patches in the LMC.  They find that
extinction largely follows a thin disk with a FWHM of about $100-200$pc,
and $\left<E(B-V)\right>=0.2$mag., averaged over the whole LMC including
the extinction by the Galactic foreground.  The internal extinction
(discounting Galactic foreground) in the optical U band
\begin{equation}
\left< A_U \right> = 4.72 \left<E(B-V)\right> = 0.6 {\rm mag}.
\end{equation}
Stronger absorption is expected in the ultra-violet.

While dust distribution in the LMC and the Galactic foreground is
known to be very clumpy and extinction is patchy, there is
surprisingly little variation on sub-pc scale.  Harris et al. find
strong variations of reddenings among members of OB associations
within a few {\it arcmins} of each other in the LMC (their Figure 13).
About 5\% of the lines of sight have low-extinction ``holes''.
Typical size of the dust patches is between $1'$ to $10'$; $1'$ is about
$15$pc at the LMC's distance.

The Galactic foreground extinction towards the LMC has also been
mapped out by Oestreicher et al. (1995) with foreground stars.  They
find reddening varies from 0 to $E(B-V)=0.15$mag. across the surface
of the LMC with a mean at $\left<E(B-V)\right>=0.06$mag..
Nevertheless the dust patches appear mostly larger than 30', which
translates to about 2pc for a line of sight path of 200pc in the thin
layer of Galactic dust.  

Extinction has also been studied in great detail towards the Baade's
window ($l=1^o$, $b=-4^o$), a relatively clear field near the Galactic
center.  Stanek (1996) uses red clump giants in the Galactic bulge
from the OGLE microlensing survey to map out the extinction, which is
due to dust patches within about $2$ kpc of the Sun (Arp 1965).  The
extinction $A_V$ varies from about $1.3$mag to $2.5$mag (99\%
confidence interval) across a $40'\times 40'$ field.  While dust
patches of $10'$ are clearly visible, we find (cf.
Fig.~\ref{figures1.ps}) that the variation of $A_V$ of neighbouring
patches of $30''$ to $3'$ is rarely more than 20\%-30\%.  The only
modest variation on these scales in this line of sight means few clouds
of size $0.3$pc to $2$pc.

\section{Basic signal}

Consider the effects of placing a thin layer of dust in the mid-plane
of a relatively thicker stellar disk of the LMC
(cf. Fig.~\ref{figures2.ps}); thin stellar disk models are less
interesting because they do not provide enough microlensing events
(Gould 1995).  Here we assume a uniform stellar disk of the LMC,
much thicker than the clumpy dust layer; the thickness is 
$1000$ pc and $100$ pc respectively.
Now nearly 50\% of the LMC stars are in front of the
dust layer, hence free from reddening (always discounting reddening by
the Galactic dust layer in the solar neighbourhood unless otherwise
specified).  Another nearly 50\% of the LMC stars are at the back the
dust layer, hence reddened by some measurable amount.  In between
there is a negligible fraction of the stars reddened by an
intermediate amount.

Now suppose that all current 30 microlensing events towards the LMC
are due to machos,  the we observe about $(15\pm 4)$ lensed sources in
front of the dust layer with negligible reddening, and $(15\pm 4)$
sources behind with some measurable amount of reddening.

In comparison, if the lenses were in the LMC disk, then there would
be a higher probability, say $1-p$, of finding sources behind the dust
layer than the probability, $p$, of finding sources in front.  It is a
simple calculation to show that $p=1/8=12.5\%$ for a uniform slab
model of the stellar disk of the LMC, and $p\approx 15\%$ for a ${\rm
sech}^2$-disk.  So star-star self-lensing models would predict only
about $(4\pm 2)$ stars out of the 30 stars to be in front of the dust
layer with only the Galactic foreground reddening
(cf. Fig.~\ref{figures2.ps}).

That self-lensing models grossly under-predict the number of
microlensed stars with the Galactic foreground reddening is the main
signal that differentiate them with macho-lensing models.  A
difference at about $3\sigma$ level is expected for the current 30 or
so microlensing events.  It appears that a long-term study of the
reddening distribution of microlensed stars towards the Magellanic
Clouds can set firm limit on self-lensing and the fraction of machos
in the halo; a quantitative analysis is given in Zhao (1999c).

\section{Practical issues and solutions}

The above arguments are robust.  The key condition is that any
measurement error of the reddening is small enough to separate stars
with only Galactic foreground reddening from those with Galactic plus
LMC reddening confidently (cf. Fig.~\ref{figures2.ps}).  The arguments
apply to a variety of star and patchy extinction distributions with a
few conditions.
\begin{itemize}
\item They are insensitive to the thickness and vertical profile of
the stellar disk as long as it is much thicker than the dust layer.
This is to be examined in detail in \S5.
\item They are insensitive to the patchiness of the dust layer of the
LMC as long as the LMC dust layer has very few ``holes''; a star
behind such a hole can be confused with one in front of the hole as
far as reddening is concerned; this happens perhaps about 5\% of the
time (Harris et al.).
\item They are insensitive to the patchiness of the dust layer of the
Galaxy as long as the Galactic foreground reddening is indeed smooth
on $10'$ scale (Oestreicher et al. 1995).
\item They are insensitive to self-extinction in a very localized
dusty cocoon as long as we avoid mass-losing AGB stars and early-type
stars in star-forming regions; better choices might be Red Giant
Branch and Clump stars and late $A-F$ type bright main sequence stars
since they are generally old enough to drift away from the dusty
cocoons at their birth place.
\end{itemize}

To check the validity of these conditions, we can use random unlensed
stars in the immediate neighbourhood of the lensed stars to map out
the dust patches in the LMC and Galactic foreground.  Polarization
maps or existing HI and CO maps of the LMC would also be helpful for
this purpose.  This way we can identify and stay away from regions
with unmeasurably low extinction.  We should exclude microlensing
candidates which happen to fall in low-extinction holes with
unmeasurable difference between stars in front and stars behind the
LMC dust layer.  We can then apply the Galactic foreground subtraction
to individual stars in the remaining sample.  The reddening
distribution of microlensed stars can then be analyzed for signs of
deficiency of ``reddening-free'' stars, an indication of
self-lensing.

Existing photometry of the microlensing survey fields are typically in
one or two broad passbands.  This is generally not enough for accurate
determination of reddening.  Reddening can be determined by
constructing reddening-free indices with photometry of three to seven
broad bands, or with low resolution spectroscopy; e.g., Terndrup et
al. (1995) show that reddening towards the Baade's window of the
Galactic bulge can be derived from the $H_\beta$ index.  Typical
accuracy is about $0.02-0.05$ mag. in $E(B-V)$ with these methods.

A practical definition of low-extinction holes might be regions with
LMC internal extinction $E(B-V) \le 0.05$mag.; these regions cover
perhaps on the order 10\% of the surface of the LMC.  Harris et
al. show that the reddening of individual OB stars can be determined
with UBVI photometry to about $\sigma(B-V)=0.04$mag., or about 30\% of
the average internal extinction of the LMC disk $E(B-V)=0.13$mag.

A way to reduce variation is to select random unlensed stars as close
to the microlensing line of sight as possible.  This way they are
likely to share the same dust patch.  The dust maps of Oestreicher et
al. (1995) suggest that Galactic foreground extinction is likely
smooth on $10'$ scale or smaller, and can be subtracted out
accurately.  For the dust in the LMC, it appears safe to work within
small patches of the sky of $4''$ scale, which corresponds to about
$1$pc in the LMC, and less than $0.004$pc in the solar neighbourhood.
At these scales variations of reddening are likely at 10\%-20\% level
among stars behind the dust layer (cf. Fig.~\ref{figures1.ps}).  Such
low-level variations would hardly affect our results since there would
be little chance of mis-classifying a star at the back of the LMC dust
layer as a star in front of the layer, even after allowing for
measurement errors at 30\% level (cf. Fig.~\ref{figures2.ps}).  It is
challenging to find enough bright unblended LMC stars from the ground
in such a tiny $4''\times 4''$ patch of the sky, though.

\section{Reddening vs. lens location}

For the clarity of the argument we adopt a set of simple models
for the density distributions of the dust $\nu_d(D)$, the lenses 
$\nu_l(D)$ and the stars $\nu_*(D)$: they
are distributed in three uniform layers with a width $w_d$, $W_*$ and $W_*$
and a mean distance $D_{\rm LMC}$, $\left<D_l\right>$, 
$\left<D_*\right>$.  We compute
the excess reddening of the microlensed star 
\beq\label{xi}
\xi \equiv {\left<A_s\right> \over \left<A_u\right>}-1
\eeq
where
\bey
\left<A_u\right> &=& {\int_0^\infty\!\! A(D) P(D)dD \over
\int_0^\infty\!\! P(D)dD},\\
\left<A_s\right> &=& {\int_0^\infty\!\! A(D_s) \tau(D_s) P(D_s)dD_s \over
\int_0^\infty\!\! \tau(D_s) P(D_s)dD_s},
\eey
are the average dust absorptions to the unlensed LMC stars
and to the microlensed LMC source stars respectively, and
\beq
A(D) = C_1 \int_0^{D} \!\! \nu_d(D_d) dD_d
\eeq
is the absorption to a star at distance $D$, and
\beq
P(D) dD = C_2 \nu_*(D) D^2 dD
\eeq
is the probability of locating a star
at distance $D$ to $D+dD$, and
\beq
\tau(D_s) = C_3 \int_0^{D_s} dD_l \nu_l(D_l) {(D_s-D_l) D_l \over D_s}
\eeq
is the optical depth to a source star at distance $D_s$.

Fig.~\ref{figures3.ps} shows the excess reddening $\xi$
as a function of the location of the lenses, 
here the renormalized typical lens distance 
$\left<D_l\right>/\left<D_*\right>$.
{\it Varying the thickness of the dust layer between
$100{\rm pc} \le w_d \le 200{\rm pc}$ 
barely makes any difference as long as the dust layer is 
thinner than the stellar disk}; 
the effect is marginally visible only for the thin disk model.
The excess reddening is a constant 70\% for purely self-lensing models 
($\left<D_l\right>/\left<D_*\right>=1$)
insensitive to the exact values of $W_*$ and $w_d$ as long as $W_* \gg w_d$.
The prediction is somewhat sensitive to
a plausible small offset between an unviralized stellar disk 
and the dust layer $\Delta_* \equiv \left<D_*\right>-D_{\rm LMC}$;
the excess reddening becomes even stronger if the stellar population 
of the LMC disk is shifted slightly closer to us than the dust layer.
In general, the excess reddening is at 1\% level if the lens population is
in the halo (the macho zone), 
and above 40\% if the lens population coincides with the stellar disk
of the LMC.  So the two scenarios are distinguishable
if we can control patchiness of reddening and measurement error
to better than 20\% level.

\section{Conclusion and strategy for observations}

In summary, we have studied effects of dust layer in the LMC on the
microlensing events.  We find that self-lensing models of the LMC draw
preferentially sources behind the dust layer of the LMC, and hence can
be distinguished from the macho-lensing models once the reddening by
dust is measured.  The effect is insensitive to the exact thickness of
the dust layer and the stellar disk (cf. Fig.~\ref{figures3.ps}).  The
deficiency of reddening-free microlensed stars is likely a robust
discriminator of the two types of lensing scenarios.

\begin{figure}
\epsfxsize=15cm \centerline{\epsfbox{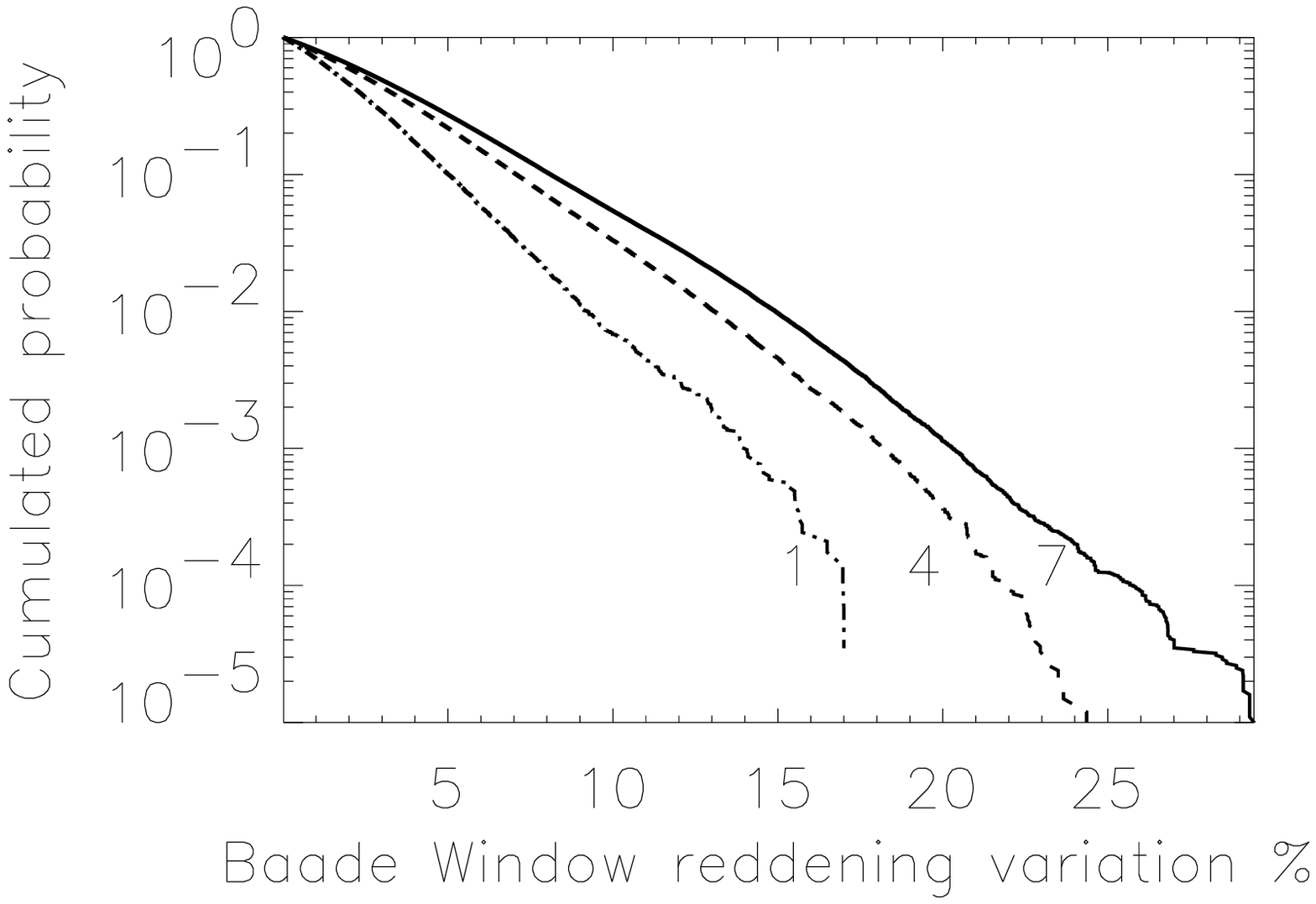}}
\caption{
Cumulated distribution of pixel-to-pixel variations 
in Stanek's reddening map of Baade's window (a 
$40' \times 40'$ field at $l=1^o$, $b=-4^o$).  
By pixel-to-pixel we mean 
any two pixels within 1, 4 or 7 pixels (solid, dashed, and dash-dotted lines)
of each other.  Each pixel is $30''\times 30''$, or about $0.3$pc in
linear size if the dust clouds are at $2$kpc.
}
\label{figures1.ps}
\end{figure}

\begin{figure}
\epsfysize=10cm \centerline{\epsfbox{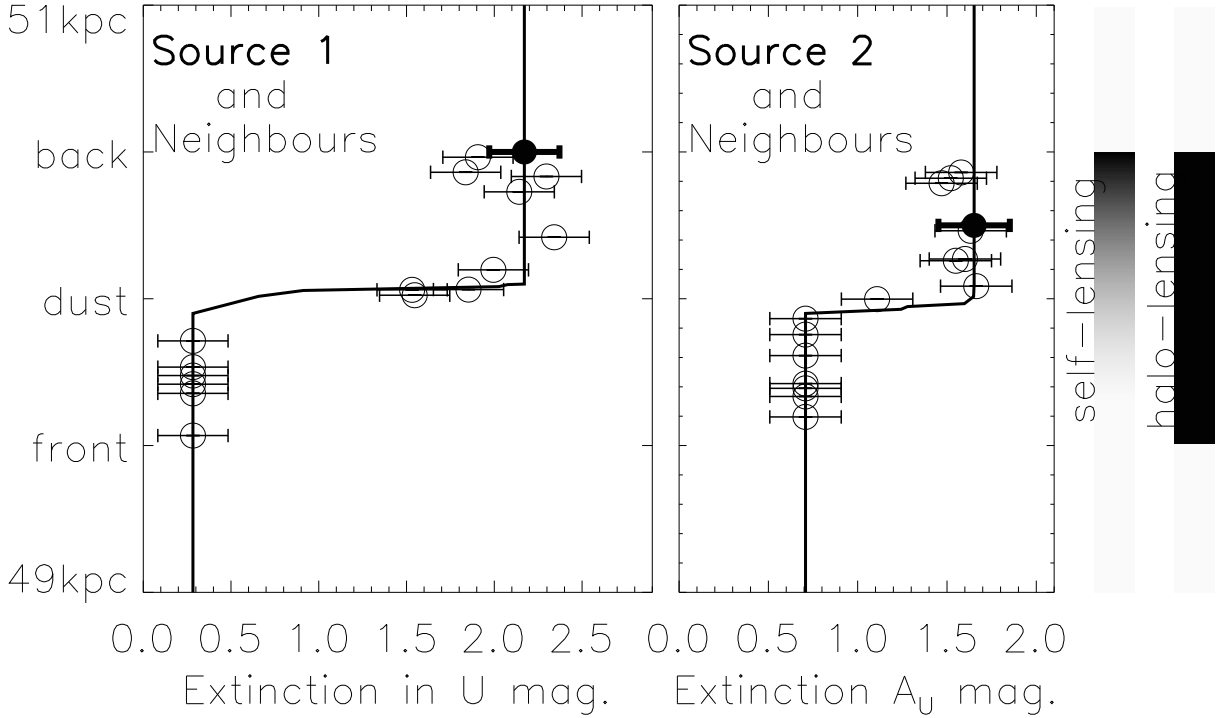}}
\caption{ {\it The two panels} show the simulated $U$-band extinction
$A_U$ as a function of line of sight distance (solid lines) 
for random LMC stars (open circles) 
in two small $4''\times 4''$ patches of the sky
centered on two hypothetical microlensed sources (filled circles) in
two widely different fields; $4''=1$pc at the LMC.  We assume a
generous measurement error of $0.2$mag in $A_U$, and 
a variation at 20\% level of the patchy internal extinction of the LMC
and no variation of the Galactic foreground within a $4''\times 4''$ patch, 
but factor-of-two variations of extinctions from patches to patches.
Note that {\it we can often classify
stars as in front of or behind the dust layer from the reddening
distribution}; it can be more difficult for low extinction regions of
the LMC (here source 2) in case of increased errors.  
{\it The two vertical bars} show the predicted distance distributions of
microlensed sources in a halo-lensing model and a self-lensing model;
the darkest region is the most probable location of the source.
Note that {\it while the macho-lensing model would draw sources with a
distribution as the unlensed stars, the self-lensing model strongly
disfavors sources in front of the dust layer with lower than average
reddening.}  }
\label{figures2.ps}
\end{figure}

\begin{figure}
\epsfysize=15cm \centerline{\epsfbox{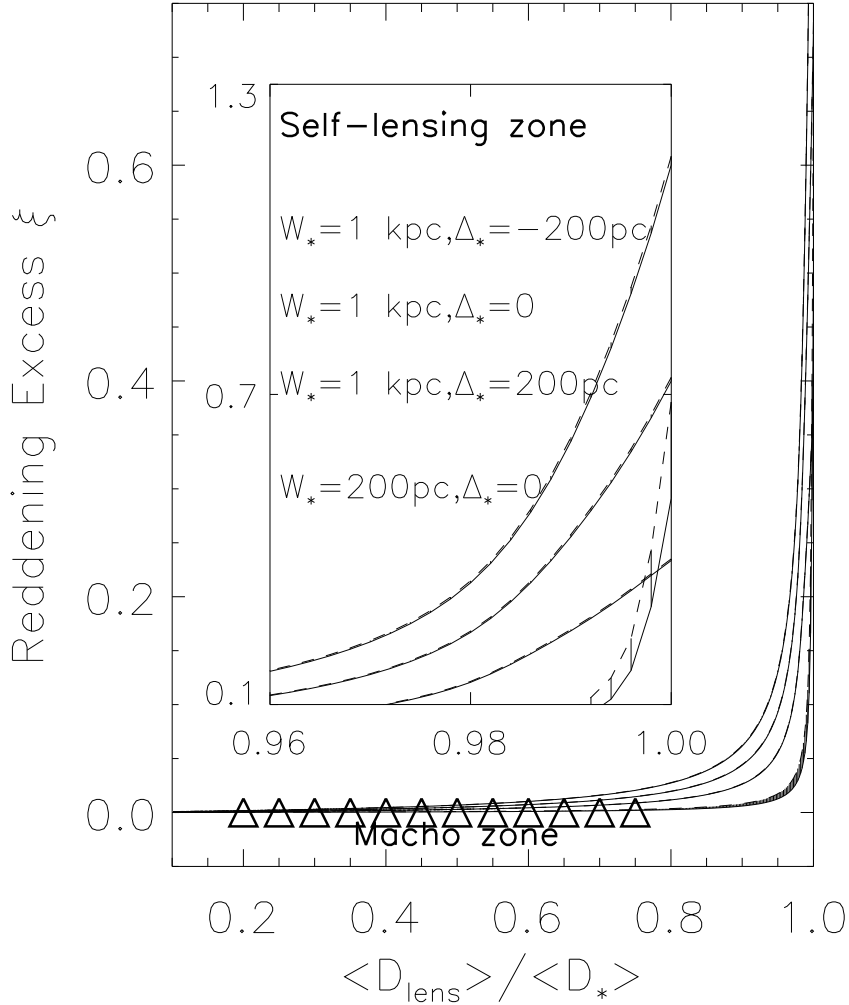}}
\caption{
The excess reddening of microlensed stars $\xi$ (cf. eq.~\ref{xi})
as an indicator of the lens location.
Foreground stars or machos would typically be at distance 
$(0.2-0.8) \left<D_*\right> =(10-40)$kpc and predict insignificant excess.  
LMC stellar lenses would be at $\left<D_l\right>=\left<D_*\right> \sim 50$kpc
in average, and predict an excess between 40\%-120\%.
Models are calculated for uniform stellar disks of width
$W_*$ from $1$kpc to $200$pc, which might be offset from the dust layer
by $\Delta_* \equiv \left<D_*\right>-D_{\rm LMC}=\pm 200$pc.
Each of the four ``curves'', 
as labeled by their parameters $W_*$ and $\Delta_*$
(in the order from top to bottom in the inset), is
actually a shaded region with the barely distinguishable 
upper and lower dashed boundaries corresponding to
models with a uniform dust layer of width 100pc and 200pc respectively.
}
\label{figures3.ps}
\end{figure}

The clumpiness of the dust, together with the fairly large error of
reddening vector derived from broad-band photometry, can lead to a
large scatter in the relation between reddening and line of sight
depth.  The finer structure of the patchy extinction in the LMC remains
to be studied as well since previous reddening maps of the LMC, e.g.,
Harris et al.'s map from the sparsely distributed luminous OB stars,
are limitted to a spatial sampling of the order $1'$ (or 10 pc).  The
trick here is to work only in small patches of the sky of $4''\times
4''$ scale, so that stars likely share the same patch of dust cloud
(cf. Fig.~\ref{figures1.ps}).  For example, the amounts of extinction
by the Galactic foreground and the LMC change wildly from one
microlensing line of sight to another (cf. Fig.~\ref{figures2.ps}),
but within each small patches the extinction is well-correlated with
the line of sight distance, and there is little ambiguity to classify
a star as in front of or behind the dust layer if we can measure the
reddening.  Another trick is to obtain as many lines of sight as
possible to beat down all variations (20\%-100\% due to patchy
extinction, and 30\%-100\% due to measurement error) by a factor
$1/\sqrt{N}$, where $N$ is the number of microlensing lines of sight.
We can differentiate the halo-lensing models from self-lensing models
at $3\sigma$ level with the reddening distribution of current 30
microlensed stars and their neighbours.

It would take about a few nights with a 2.5m telescope at a good
seeing site to obtain accurate (1\%) photometry in several broad or
narrow bands for the present sample of microlensed stars and
neighbouring stars in the $V=20-21$mag range.  Seeing is critical to
reach fainter stars.  Nevertheless, there are two general problems of
ground studies.  From the ground the fainter ($V>20$mag.) LMC stars
are often blended in the $1''-2''$ seeing disk, which results in
spurious colors, hence unphysical reddening.  A $4''\times 4''$ patch
of the LMC might show only a handful of $V=20$ mag stars but contain 
hundreds of fainter objects at the resolution limit of 
the Hubble Space Telescope (HST).  Second it is
difficult to access the ultra-violet band from the ground, which is
the most sensitive band for measuring dust absorption.  For these
reasons, photometry or spectroscopy in the ultra-violet from HST is
desirable for getting an unambiguous answer.

\vskip 1cm

The author thanks Paul Hodge, Puraga Guhathakurta, David Spergel, Tim
de Zeeuw for encouragements, Walter Jaffe and Frank Israel for
enlightening discussions and Bryan Miller specially for many helpful
comments on the presentation.

\vfill \eject

{}


\label{lastpage}

\end{document}